\documentclass[a4paper,11pt]{article}
\pdfoutput=1 

\usepackage{jinstpub} 
\usepackage{lineno,hyperref}
\usepackage{color,soul}
\usepackage{comment}
\usepackage{float}
\usepackage{graphicx}
\usepackage{booktabs}
\usepackage{subcaption}
\usepackage{amsmath}
\usepackage{url}
\usepackage{xcolor}
\usepackage{natbib}

\title{Measurement of Proton Quenching in a Plastic
Scintillator Detector}

\author[a]{Connor Awe,}
 \author[a]{Phillip Barbeau,}
 \author[b]{Alireza Haghighat,}
 \author[a]{Sam Hedges,}
  \author[a]{Tyler Johnson,}
\author[c,1]{Shengchao Li,\note{Corresponding author.}}
\author[c]{Jonathan M. Link,}
\author[b]{Valerio Mascolino,}
 \author[a]{Jay Runge,}
\author[c,d]{Jacob Steenis,}
\author[c]{Tulasi Subedi}
  \author[c]{and Keegan Walkup}

\affiliation[a]{Department of Physics, Duke University,
 and Triangle Universities Nuclear Laboratories, \\
 Durham, NC, USA}
 \affiliation[b]{Nuclear Engineering Program,   Virginia Tech, \\
 Northern Virginia Center, VA, USA}
\affiliation[c]{Center for Neutrino Physics, Department of Physics, Virginia Tech, \\
Blacksburg, VA, USA}
\affiliation[d]{Department of Physics, Grinnell College,\\
Grinnell, IA, USA}

\emailAdd{scli@vt.edu}

\abstract{The non-linear energy response of the plastic scintillator EJ-260 is measured with the MicroCHANDLER detector, using neutron beams of energy 5 to 27 MeV at the Triangle Universities Nuclear Laboratory.  The first and second order Birks' constants are extracted from the data, and found to be $k_B = (8.70 \pm 0.93)\times 10^{-3}\ {\rm g/cm^2/MeV}$ and $k_C = (1.42 \pm 1.00) \times 10^{-5}\ {\rm (g/cm^2/MeV)^2}$.  This result covers a unique energy range that is of direct relevance for fast neutron backgrounds in reactor inverse beta decay detectors.  These measurements will improve the energy non-linearity modeling of plastic scintillator detectors.  In particular, the updated energy response model will lead to an improvement of fast neutron modeling for detectors based on the CHANDLER reactor neutrino detector technology. }

\keywords{Scintillators and scintillating fibres and light guides; Neutrino detectors; Neutron detectors (cold, thermal, fast neutrons); Simulation methods and programs}

\arxivnumber{2011.11103} 

\begin{document}
\maketitle
\flushbottom

\bibliographystyle{unsrt}

\section{Introduction}
This paper presents measurements of the scintillation light quenching of protons, across a range of energies, for Eljen Technology's wavelength-shifting plastic scintillator, EJ-260, which is used in the CHANDLER reactor neutrino detector technology.  There have been light output measurements made on similar materials~\cite{Pozzi:2004dn,Weldon:2020,8933026,Laplace:2020mfy}, but to our knowledge, this is the first such measurement on EJ-260.  In addition, this measurement covers a wider energy range than previous measurements, extending from 5~MeV up to 27~MeV, which is relevant for the cosmic-ray fast neutron backgrounds that dominate in surface-level reactor neutrino detection. 

\subsection{Description of the CHANDLER Technology}
CHANDLER is a reactor antineutrino detection technology based on solid plastic scintillator designed to reject the high-background environment at the surface level~\cite{CHANDLER_Patent}.  It consists of layers of plastic scintillating cubes arrayed to form a Raghavan optical lattice (ROL)\@. The ROL transports light to the surface of the detector by total internal reflection along the cube rows and columns, where it is collected by photomultiplier tubes (PMTs)\@.  In the ROL, an event can be localized to the cube that sits at the intersection of an active row and column.  The cube layers are separated by thin sheets of $^6$Li-loaded ZnS scintillator, which are used for neutron detection.  The plastic scintillator is doped with a wavelength-shifting compound which absorbs the blue light emitted by the ZnS scintillator and re-emits it so that it can also be transported by total internal reflection.  The longer scintillation decay time of the ZnS scintillator (200\,ns) relative to the plastic scintillator (10\,ns) enables neutron identification through pulse shape discrimination.

In a CHANDLER detector, antineutrinos are observed via the inverse beta decay (IBD) process in which an electron antineutrino interacts with a proton in the plastic scintillator, producing a neutron and a positron.  The positron deposits its kinetic energy in the plastic scintillator and annihilates with an electron, producing two 511\,keV gammas.  A prompt signal is produced by the positron and the annihilation gammas, which Compton scatter in neighboring cells.  The neutron thermalizes and is captured by the $^6$Li in the sheets, converting to an alpha and a triton. These charged particles deposit their energy in the ZnS scintillator producing a delayed signal.  The signature of an IBD event is the coincidence of prompt and delayed signals in both space and time.  The largest background to this process comes from cosmic-ray fast neutrons, which can scatter off a proton in the detector, creating an energetic recoil, that mimics the positron from IBD\@.  The neutron can subsequently thermalize and capture on $^6$Li, completing the prompt/delayed coincident pair.

\subsection{Importance of this Quenching Factor Measurement}
On average, energetic fast neutrons will have a slightly larger spatial separation than IBD events, but the temporal separation is the same.  As a result, the elimination of fast neutron backgrounds relies on a clear understanding of the fast neutron's behaviour in the detector.  This model is complicated by quenching effects in the plastic scintillator, which give it a non-linear response to deposited energy.   Previous studies have measured the energy response of plastic scintillator to neutrons with energies less than 5~MeV~\citep{Pozzi:2004dn,8933026,Laplace:2020mfy} and less than 10~MeV~\citep{Weldon:2020}, which are mostly below the relevant energy for cosmic-ray fast neutrons in our detector, particularly in light of these measurements which correspond to an electron equivalent energy response of 1.8 MeVee or less for a 5~MeV neutron.  In this paper, we establish a model of the proton quenching factor based on 16 distinct measurements with neutron energies from 5 to 27 MeV.  

\section{Scintillation Light Quenching}
The dependence of the scintillation response on the particle type, particularly the divergence of nuclear recoils from proportionality, was first discussed by Birks~\citep{Birks:1964zz}.  Birks' law is an empirical model used to describe these non-linear quenching effects.  Here a generalized model~\citep{Chou:1952jqv} was adopted: 
\begin{equation}\label{eqn:birks1}
    \frac{dL}{dx}=S\frac{\frac{dE}{dx}}{1+k_B(\frac{dE}{dx})+k_C(\frac{dE}{dx})^2} ,
\end{equation}
where $dL/dx$ is the light yield per unit length, S is the scintillation efficiency, and $k_B$, $k_C$ are the first and second order Birks' constants.  The energy loss per unit length in the medium, $dE/dx$, is a strong function of the proton energy. 

The quenching factor is the ratio between the observed scintillation energy and the true proton recoil energy:
\begin{align}\label{eqn:birks2}
    QF(k_B,k_C,E_p)&=\frac{E_{vis}(k_B,k_C,E_p)}{E_p}\\
    &=\frac{1}{E_p} \int_0^{E_p}\frac{dE}{1+k_B(\frac{dE}{dx})+k_C(\frac{dE}{dx})^2}  \label{eqn:birks3}
\end{align}
where $E_p$ is the true proton recoil energy, $E_{vis}$ is the visible scintillation energy, proportional to the $dL/dx$ in Eq.~(\ref{eqn:birks1}), and the Birks' constants, $k_B$ and $k_C$, are properties of the scintillator material.

\section{Experimental Setup}
To measure the quenching factor for recoil protons in the CHANDLER scintillator, the MicroCHANDLER prototype detector was exposed to a neutron beam at the Triangle University Nuclear Laboratory (TUNL) in Durham, North Carolina.  

\subsection{The MicroCHANDLER Detector}
MicroCHANDLER is a smaller version of the MiniCHANDLER detector that was used to detect reactor antineutrinos at the North Anna Nuclear Generating station during a deployment in 2017~\cite{Haghighat:2018mve}.  The two detectors share the same cube and sheet structure, with MiniCHANDLER being an $8\!\times\!8\!\times\!5$ cube array, while MicroCHANDLER is a $3\!\times\!3\!\times\!3$ array.  Both detectors use the same scintillator materials.  The plastic scintillator is based on polyvinyltoluene (PVT) and is doped with a wavelength-shifting compound.  It is sold commercially by Eljen Technology as EJ-260.  It is this scintillator that is the subject of the quenching factor measurement presented in this paper. 

MicroCHANDLER is read out on two sides.  The opposing sides are covered with aluminized Mylar sheets to reflect light back towards the PMTs.  The detector enclosure has been rendered light-tight.  The PMT signals are amplified and shaped in 25\,ns, and feed into a CAEN DT5740 waveform digitizer with a 12-bit ADC and 62.5\,MHz sampling rate.

\begin{figure}
    \centering
    \includegraphics[width = .5\textwidth]{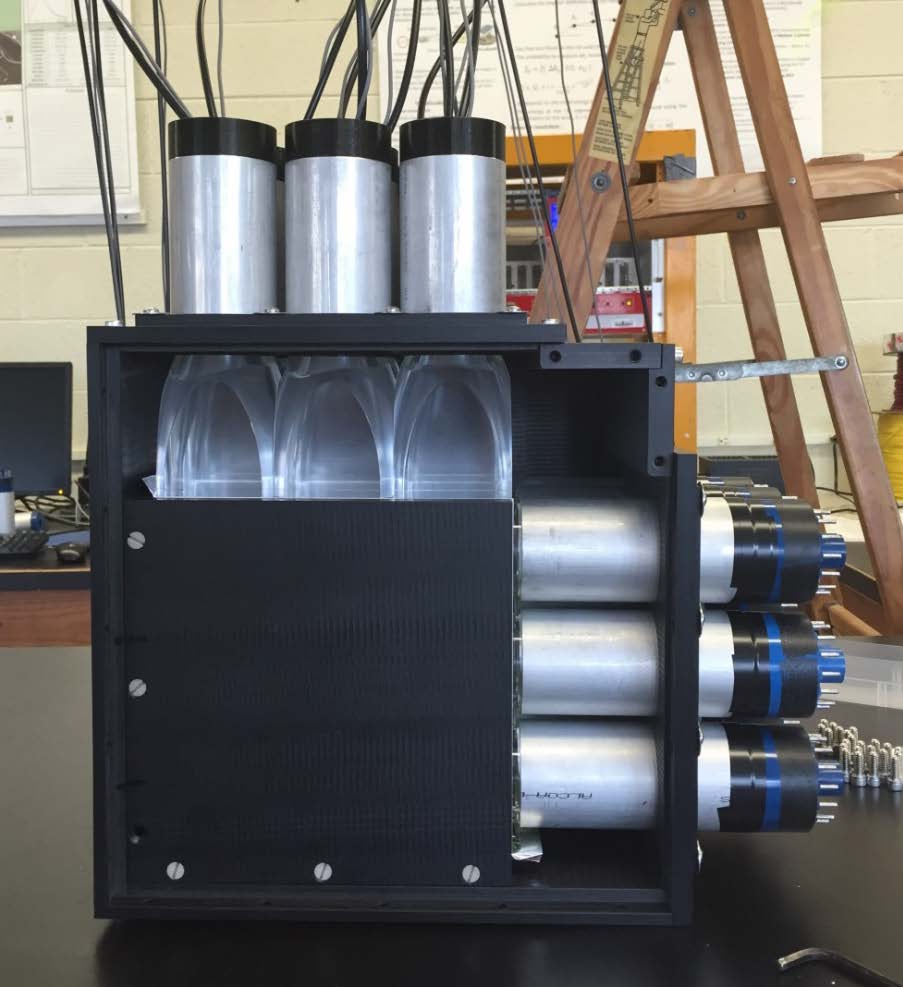}
    \caption{The hybrid MicroCHANDLER detector with old PMTs on the right and new PMTs with light guides on top.}
    \label{fig:uCHANDLER}
\end{figure}

\begin{figure}
\centering
\begin{subfigure}{0.49\linewidth}
    \includegraphics[width=\linewidth]{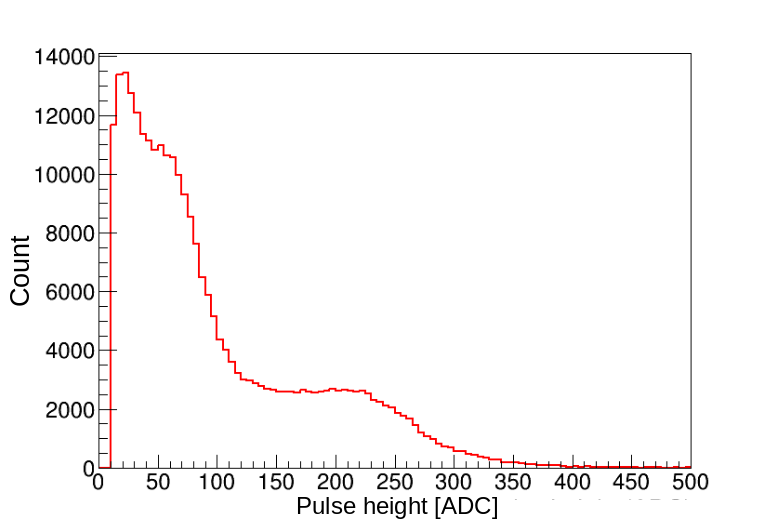}
\end{subfigure}%
\begin{subfigure}{0.49\linewidth}
    \includegraphics[width=\linewidth]{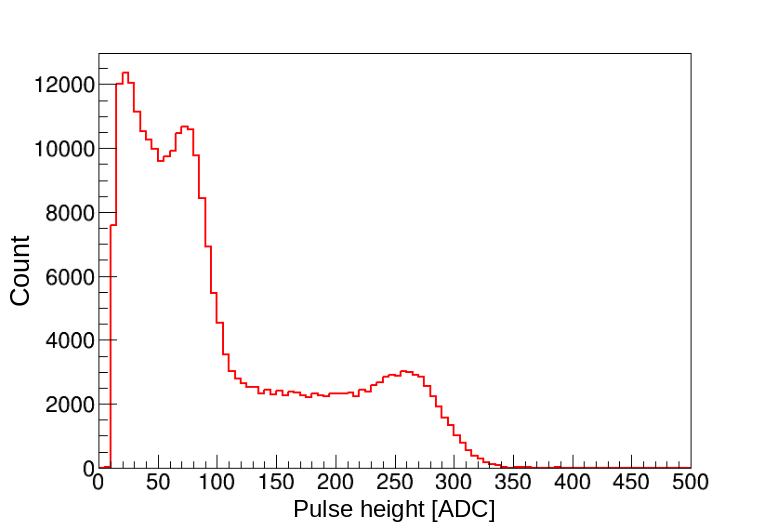}
\end{subfigure}
    \caption{The observed pulse height spectra in the hybrid MicroCHANDLER detector after exposure to a $^{22}$Na gamma source with old PMTs (right) and new PMTs with light guides (left). }
    \label{fig:old_new_CE}
\end{figure}

MicroCHANDLER serves as a test bed for upgrades to future detectors. It includes new PMTs (Hamamatsu R6231-100) and compound parabolic light guides, which were not used in the 2017 version of MiniCHANDLER.  Fig.~\ref{fig:uCHANDLER}, shows MicroCHANDLER in a hybrid state with one side instrumented with the old PMTs (Amperex XP2202) and the other side instrumented with the new PMTs and light guides.  It was used in this hybrid-state to make a comparison of the energy resolutions of the old and new configurations.  A $^{22}$Na gamma source, producing 511~keV and 1274~keV gammas, was placed on top of the detector, and the resulting waveforms were recorded.  The observed pulse height spectra from the old and new configurations are plotted in Fig.~\ref{fig:old_new_CE}.  In the new configuration the 1274~keV Compton edge shows a factor of two improvement in energy resolution compared to the old configuration.  Also, in the old configuration, the 511~keV Compton edge is just a shoulder on the low-energy pedestal, while in the new configuration it is a distinct feature.      
{}
In this experiment, all of MicroCHANDLER's 18 channels are of the the new configuration.  

\subsection{The Neutron Beam}
\label{sec:beam}
Two measurement campaigns were carried out using TUNL's tandem Van de Graff accelerator facility.  The accelerator has a terminal bias range of 0 to 10~MV and utilizes a direct extraction negative ion source to supply pulsed, unpolarized D$^-$ or H$^-$ ions.  In a tandem Van de Graff accelerator, the negative ions have their electrons stripped and are sent back through the voltage gap resulting in a beam energy of twice the terminal voltage, or 0 to 20~MeV\@.  The beam operates at a maximum frequency of 2.5~MHz corresponding to ion bunches separated by 400~ns.  Lower frequencies can be achieved by throwing away bunches.  Neutrons were produced by directing a deuteron beam onto a tritiated target, producing nearly monochromatic neutrons via the D-T reaction.  Additionally, deuterium embedded in the target during previous experiments resulted in a second population of lower energy neutrons produced by the D-D reaction.  By scanning the energy of the incident deuteron beam, it is possible to produce neutron beams with a range of energies.  The neutron beam energy is characterized at each step using both a time-of-flight (ToF) measurement from the target to MicroCHANDLER and a detailed simulation described in Sec.~\ref{Sec:MCNP}. The corresponding proton recoil spectrum extends from the neutron beam energy down to zero.  For a given beam energy, the quenching factor is measured from a fit to this spectrum. An independent calculation of the ToF using a standalone liquid scintillator detector was attempted, but was unusable due to large uncertainty bands resulting from the limited standoff distance available in the target room. 

\subsection{Data Collection}
\label{sec:data-taking}
In this experiment the MicroCHANDLER detector was positioned directly downstream from the deuteron beam line.  The detector was surrounded by panels of borated polyethylene to reduce the asynchronous backgrounds from thermal capture.  The detector was triggered internally by a simple threshold trigger of 50 ADCs, or roughly 0.25 MeVee.  Each trigger initiates a readout of a 129 sample waveform from each of the detector's 18 channels, and from the beam pulse monitor (BPM)\@.  The BPM is timed to fall in the readout window for all beam correlated events and it is used as the reference time when calculating ToF for beam events.  The tandem beam was tuned to eight distinct terminal voltages corresponding to a total of 16 distinct neutron energies between 5 and 27~MeV.  Periodic measurements were also taken with a stand-alone ToF detector to monitor beam energy.

\subsection{Calibration}
During each night of the run, an 8~hour \textit{in-situ} muon calibration was performed, using a higher trigger threshold.  This served two purposes: 1) to calibrate the PMT gain in each channel, and 2) to calibrate the detector's energy response for the event reconstruction.   In order to determine the PMT gain, a Landau function was fitted to the muon data to track gain drift in the detector's 18 PMTs.  Only one channel experienced a gain drift of more than 2\% throughout the four day campaign, and that gain drift was 4\%.   

The detector's energy response was characterized using a subset of muon data selected by requiring three consecutive hits in the same vertical column of cubes.  As minimum ionizing particles with a tightly constrained track length, these vertical muons make a good fixed energy source to calibrate the the detector.  First, a GEANT4~\citep{Agostinelli:2002hh} simulation was used to get a distribution of the true cube energy depositions for muons satisfying the vertical selection.  This is fit with a Landau function to obtain the peak value (11.9~MeV).  Then the vertical muons' ADC distribution is fit, on a cube-by-cube basis, for each position in a layer.  A 9$\times$6 matrix for converting MeVee to ADC is formed, which automatically incorporates energy calibration in the event reconstruction.  This reconstruction algorithm is a modified version of the one used for MiniCHANDLER.  It implicitly includes all effects from light attenuation, scattering, and electronics cross-talk~\citep{Haghighat:2018mve}.

\section{Modeling and Simulation}
A detailed computational model of the MicroCHANDLER detector has been developed using the Monte Carlo particle transport code MCNP~\cite{goorley2012initial}\@. The model includes features of the experimental setup as it was performed at TUNL\@.  The supporting model was intended to predict and compare the spectrum of the proton recoil (i.e., neutron scattering on the scintillator's hydrogen atoms) events inside the cubes to provide a theoretical reference for assessing the quenching effect in the detector.

\subsection{The MicroCHANDLER at TUNL MCNP Model \label{Sec:MCNP}}
\begin{figure}

\includegraphics[width=\linewidth]{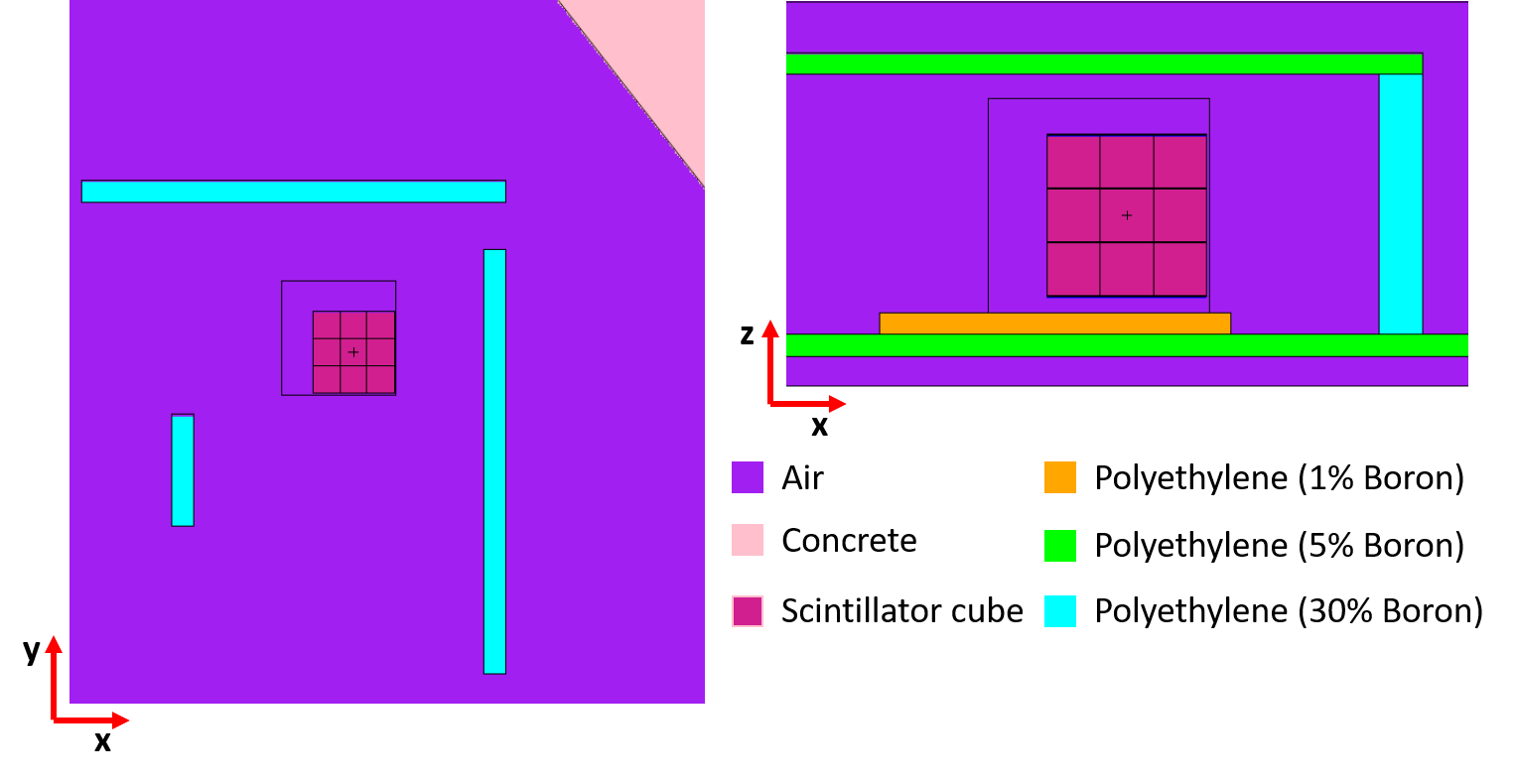}
\caption{Sections of the MicroCHANDLER MCNP model: $x$-$y$ (left) and $x$-$z$ (right) planes.}
\label{fig:ch_model}
\end{figure}

The MicroCHANDLER model consists of a $3\!\times\!3\!\times\!3$ array of scintillator cubes separated vertically by $^6$Li-loaded ZnS scintillator sheets. The detector is positioned in a neutron shield made of borated polyethylene (BP) panels that were added to reduce the thermal neutron background signal.  Some BP panels have different boron concentrations, which is taken into account in the model.  The two walls in the experimental hall that are closest to the detector are modeled as standard Portland concrete~\cite{portland}.  The dimensions of the BP shield, the distance from the beam window to the detector ($L$), and distance from the detector to the walls were all measured at the start of the run, so the ``as built'' values are accurately captured in the MCNP model.  The metallic cart that was used as a stand for the detector was not modeled, as the contribution of neutrons scattering off it and into the detector was deemed irrelevant due to the small scattering cross-sections of its components.  Detector cross sections from the MCNP model are shown in Fig.~\ref{fig:ch_model}.

\begin{table}
    \centering
    \begin{tabular}{| c | c | c |} \hline
        \textbf{Beam Terminal Voltage} & \textbf{D-D Neutron Energy} &\textbf{D-T Neutron Energy} \\
        \hline 
        0.5 MV & 5.32 MeV & 18.35 MeV \\
        1.0 MV & 6.50 MeV & 19.86 MeV\\
        1.5 MV & 7.60 MeV & 21.20 MeV \\
        2.0 MV & 8.65 MeV & 22.45 MeV \\
        2.5 MV & 9.68 MeV & 23.64 MeV \\
        3.0 MV & 10.68 MeV & 24.78 MeV \\
        3.5 MV & 11.67 MeV & 25.90 MeV \\
        4.0 MV & 12.66 MeV & 27.00 MeV \\
        \hline 
    \end{tabular}
    \caption{Neutron energies utilized in Monte Carlo simulations based on TUNL beam terminal voltages}
    \label{tab:nsource}
\end{table}

In the Monte Carlo simulation, 16 independent monochromatic neutron sources are started at the titanium tritium (TiT) target surface. The 16 beam energies ($E_{beam}$) are based on the D-T and D-D neutron energies reconstructed from the ToF measurements for the different terminal voltages utilized during the experiments (see Secs.~\ref{sec:beam} and \ref{sec:data-taking}, and Fig.~\ref{fig:srim_dE_model}). The neutron beam energies are listed in Tab.~\ref{tab:nsource}.

The neutron source is modeled as a point, given the small size of the TiT target (1.3 cm diameter) when compared to the size of the detector ($\sim$19 cm side) and the distance from the target to the front face of the detector along beam line (341.2 cm).  The angular distribution ($\theta$) of the neutron source in the model is adjusted to match the maximum acceptance angle based on the experiment geometry.  Neutrons are sampled uniformly within the allowed range.  Simple trigonometry is used to verify that the maximum acceptance angle of the detector is $\theta\approx4.73^\circ$. However, in order to include the effects of neutrons scattering off the polyethylene shield, the beam-spread in the simulation has been extended to $\theta^*=10^{\circ}$.

A MATLAB model of the kinematic interactions of the deuteron beam with the D and T atoms in the TiT target was developed as part of this work.  This model confirms that neutrons generated from the D-D and D-T reactions and falling within the acceptance range of the detector, $\theta^*$, are essentially mono-energetic (with variations $<0.1\%$) and uniformly distributed over the cosine of the acceptance angle ($\cos\theta^*$). 

In each of the MCNP runs, neutrons are followed as they interact within the cubes and the surrounding materials, such as the polyethylene shield and the concrete walls, and as they produce secondary particles like photons (from capture reactions) and protons, as recoils from scattering interactions.

The Monte Carlo calculations provide the following information:
\begin{itemize}
    \item The location and time of scattering events (in terms of geometrical coordinates and cube IDs), the energy of the recoil protons, which is deposited in the material via ionization loss.
    \item The location and time of generation of Compton scattering and photon absorption interactions, as well as the energy deposited in the detector cubes as a result of these interactions.
    \item The average cube-wise neutron interactions for 3 neutron energy groups: thermal ($E<5\,$eV), epithermal ($5\,{\rm eV}<E<1\,$ MeV), and fast ($E>1$ MeV).
    \item An energy vs.\ time heat map of the proton and photon interactions in the detector. 
\end{itemize}
From the energy deposition of protons and photons in the detector cubes, it is possible to reconstruct the theoretical spectrum (total and cube-wise) of the MicroCHANDLER detector response, as well as the relative contributions of different particle interactions. The analysis suite for the MCNP output also allows the imposition of a time filter to select events from within a certain time of the initial neutron interaction in the detector. 

In addition to the above quantities, cube-wise averages for the various neutron interactions in the detector are obtained and analyzed to understand how neutrons contribute to the detector response in different cubes based on their energy.  This will help in the subsequent design and optimization of larger CHANDLER detectors, as well as informing the topological selection criteria for discriminating between fast neutrons and IBD events based on their energy deposition patterns. 

\subsection{Monte Carlo Analysis Results}
\label{sec:mc-results}

The main function of the simulation analysis is to model the theoretical spectrum of energy deposition within the detector cubes.  This MC spectrum is then quenched and smeared for comparison to the corresponding spectrum from the 8 experimental runs (see Tab.~\ref{tab:nsource}).

\begin{figure}[t]
    \centering
    \includegraphics[width=\linewidth]{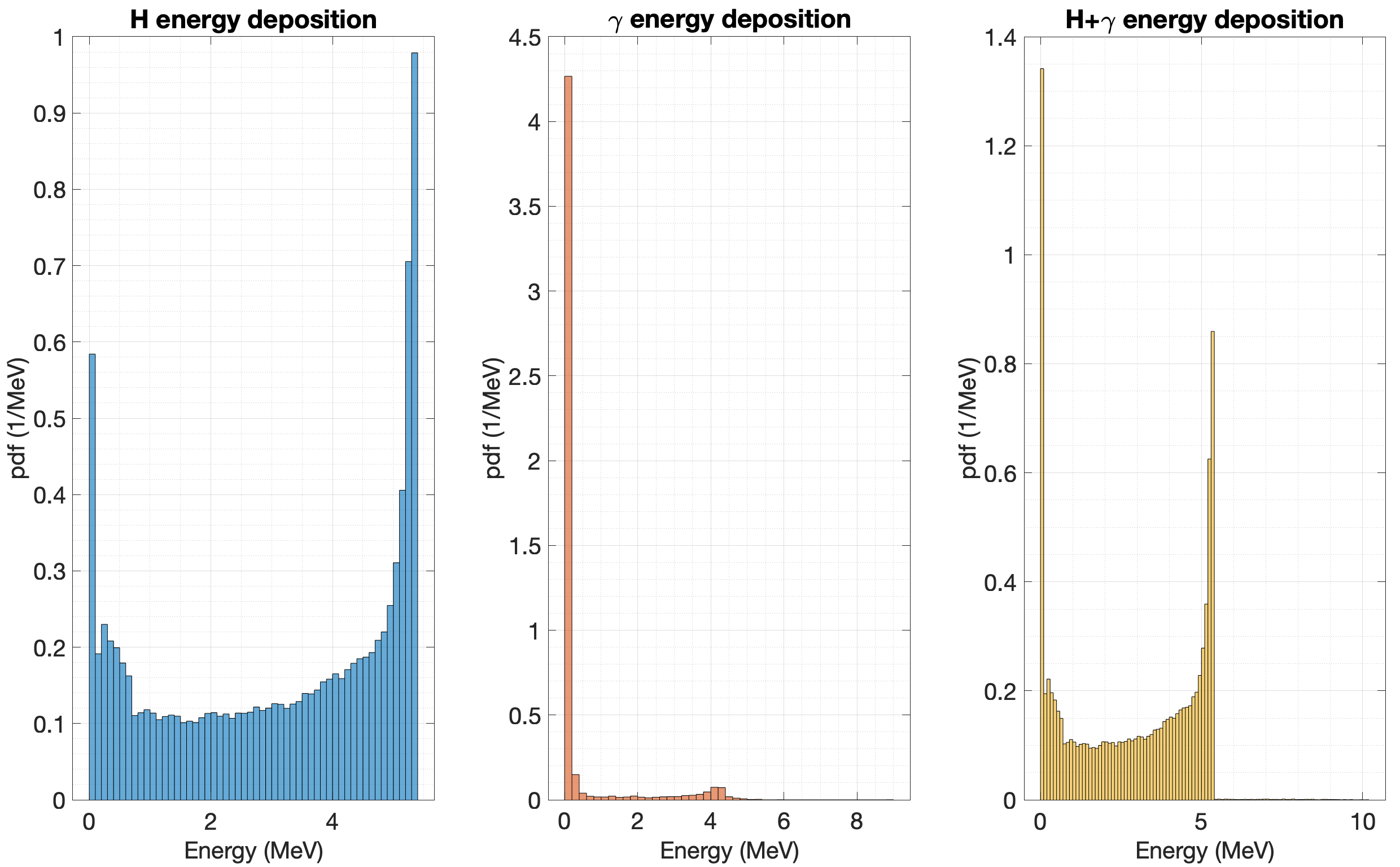}
    \caption{An example of the MCNP-calculated energy deposition spectrum for protons (left), photons (center), and total (right) summed over all the cubes of the MicroCHANDLER detector.  These plots correspond to a beam energy of 5.32~MeV with an angular spread of $10^{\circ}$.}
    \label{fig:histogram_5deg_beam1}
\end{figure}

An example of an energy deposition spectrum from the MCNP model is provided in Fig.~\ref{fig:histogram_5deg_beam1}.  A time filter is applied to the event energy: any interaction that happens more than 100~ns after the first neutron scattering in the detector is not included in the event.  This removes nearly all energy depositions from neutron capture. 

\begin{figure}[h]
    \centering

    \includegraphics[width=\linewidth]{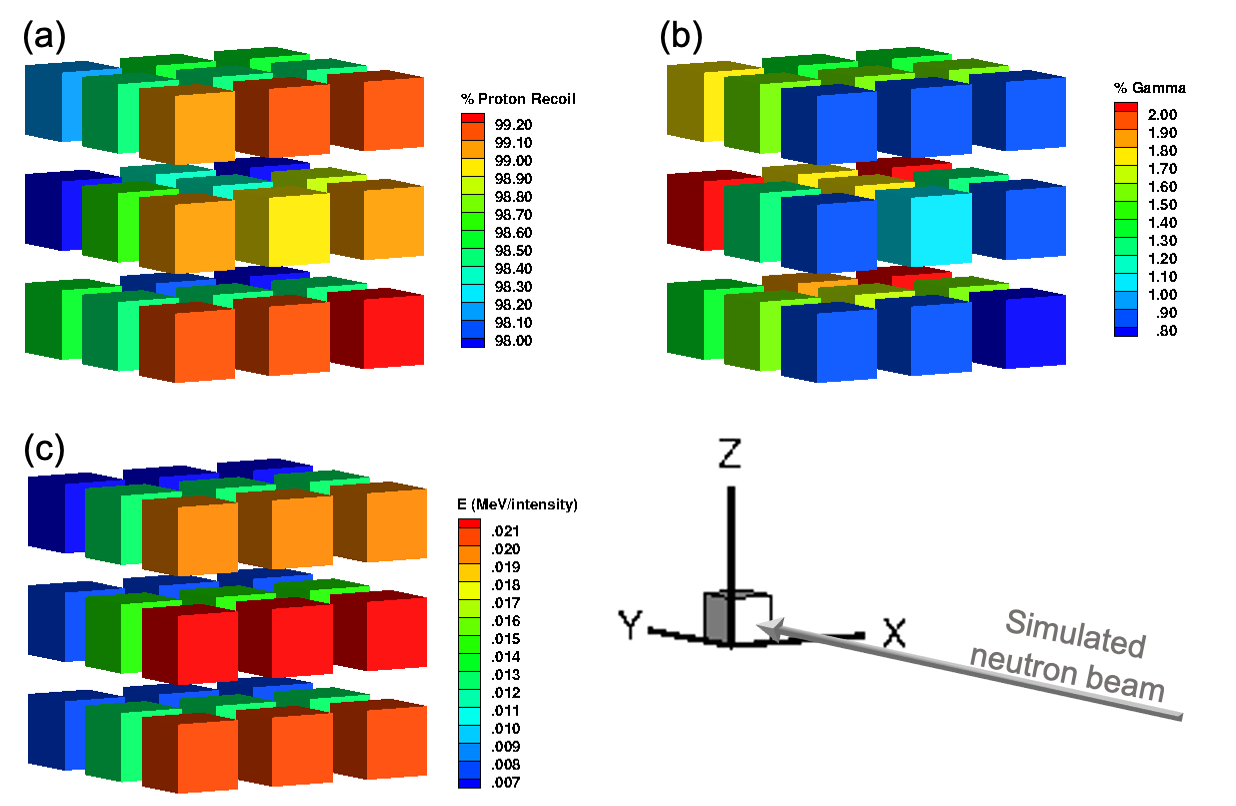}
    \caption{MCNP-calculated event-averaged fractional energy depositions for (a) protons, (b) photons, and (c) total energy depositions per simulated neutron for each cube in MicroCHANDLER.  The energies in each cube are averaged over all the events simulated for a neutron beam of 5.32~MeV energy, with an angular spread of $10^{\circ}$.  The neutron beam is directed along the positive orientation of the $y$-axis and centered around the middle cube on the $x$-$z$ plane (shown with a gray arrow).}
    \label{fig:Edep_cubes_spr5}
\end{figure}

The energy deposition distribution within the cubes depends significantly on which particle is depositing the energy (proton or photon). Figs.~\ref{fig:Edep_cubes_spr5}a\&b show the energy deposition from recoil protons and gammas for the same neutron beam energy.

Fig.~\ref{fig:Edep_cubes_spr5}c demonstrates how the energy deposition from fast neutrons in the MicroCHANDLER cubes is mostly due to proton recoils, while very little contribution comes from photons -- differing by around 2 orders of magnitude.  The calculation demonstrates that photons are only responsible for a small fraction of the energy deposition in the detector.

\begin{figure}
    \centering
    \includegraphics[width=0.6\linewidth]{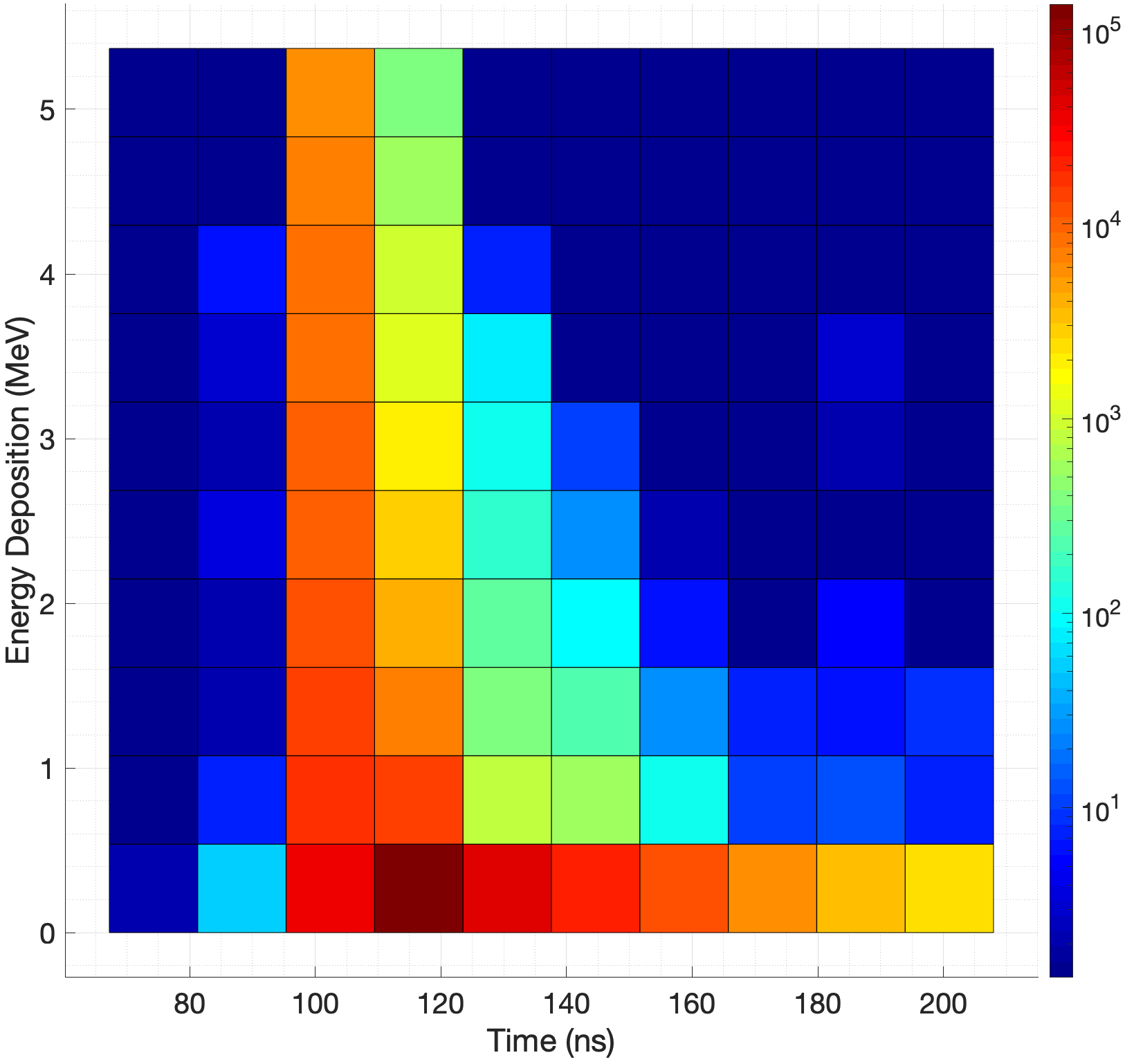}
    \caption{Energy deposition vs. time of interaction heat map in the detector.  This plot corresponds to a beam energy of 5.32~MeV with an angular spread of $10^{\circ}$.}
    \label{fig:energy_time_heatmap}
\end{figure}

In order to identify \textit{when} in time most of the interactions happen, the analysis suite produces heat maps for energy deposition vs.\ time of interaction, as shown in Fig.~\ref{fig:energy_time_heatmap} for the same neutron beam.  Note that the origin time, $t=0$, here refers to when the neutrons are emitted from the TiT target surface.
Fig.~\ref{fig:energy_time_heatmap} demonstrates that the higher energy interactions (i.e., the deposition of energy in the cubes from proton recoils) happen between 100 and 120~ns after the beam strikes the target, which is the expected ToF window for a 5.32~MeV neutron to first interact in the detector.

\section{Data Analysis}
\subsection{Time-of-Flight Energy}
\begin{figure}[h]
\centering
\includegraphics[width=0.8\linewidth]{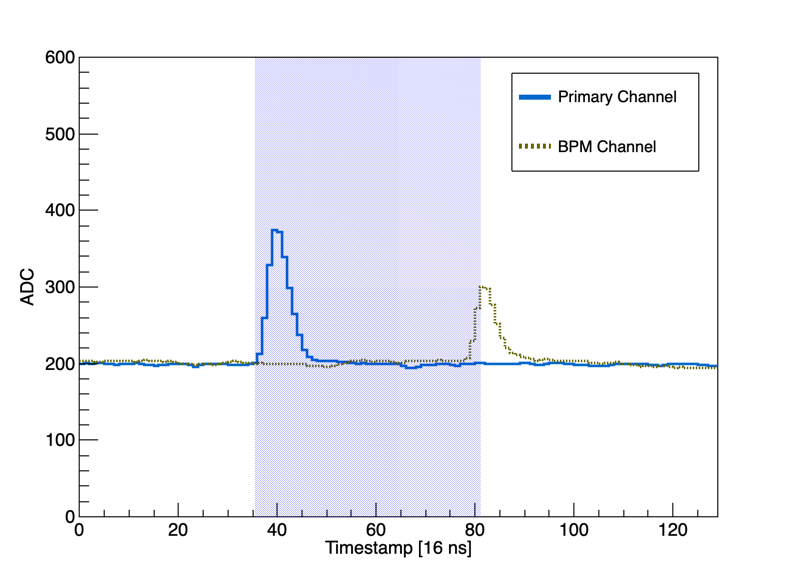}
\caption{An example of waveforms from the primary PMT (blue solid line) and the BPM (red dashed line). The blue area is the timing offset which is -729.6~ns.}
\label{fig:waveform}
\end{figure}
\begin{figure}[h]
\centering
\includegraphics[width=0.8\linewidth]{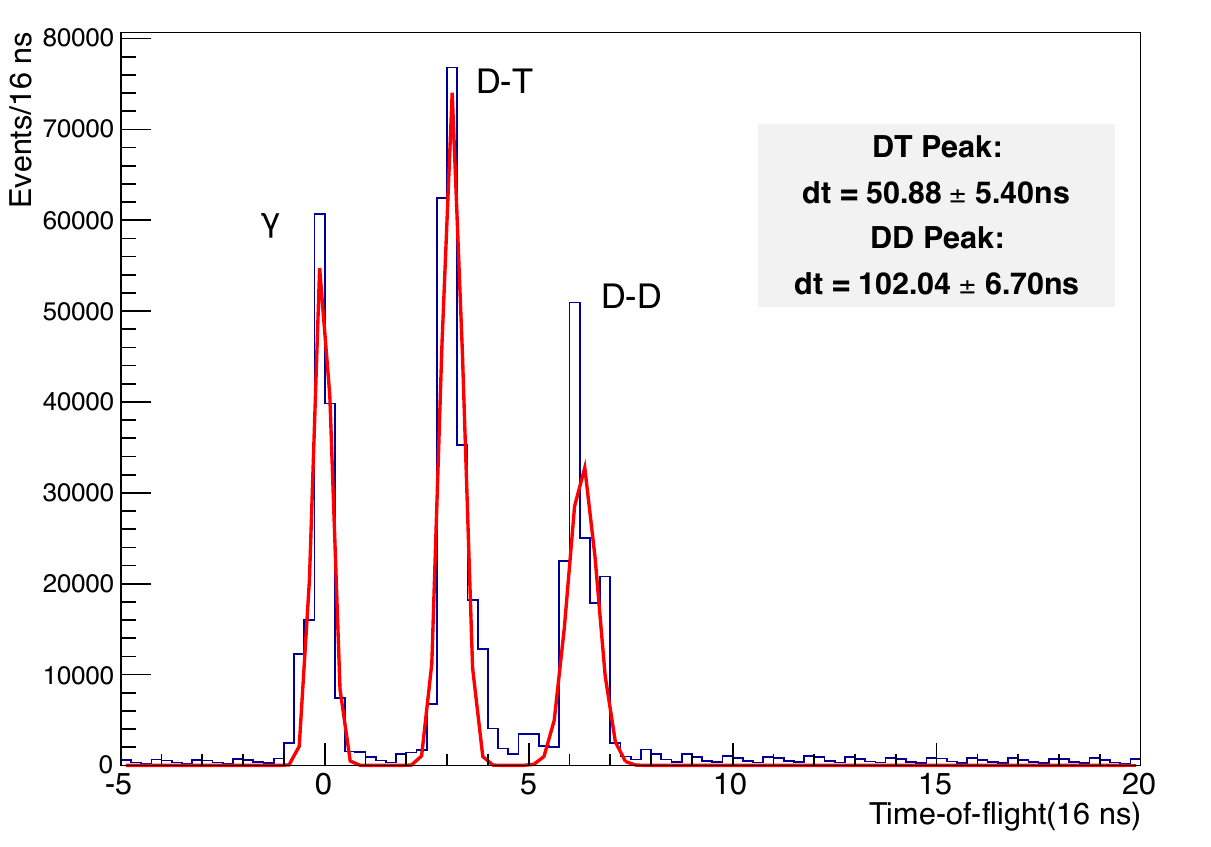}
\caption{Distribution of the ToF with the left peak aligned at zero. The left peak is the Compton scattering gammas. The middle peak is the quasi-elastic scattering neutrons from D-T reactions, and right peak from D-D reactions. The neutron peaks are well-separable in all runs. }
\label{fig:TOF}
\end{figure}

The kinetic energy of the neutrons is calculated from their ToF from the production target to the MicroCHANDLER detector.  Gammas can also be produced in the the target materials, and we use these beam gammas events to fix the ToF start time relative to the BPM.  The beam pulse in the BPM channel was adjusted with a pulse delay generator to appear in the same 2.064~$\mu$s trigger window as the beam events.  The offset between a scintillation pulse and the BPM pulse is determined by a linear interpolation of the rising edge of the PMT waveform and the peak of the BPM waveform, as shown in Fig.~\ref{fig:waveform}.  This interpolation achieves a finer timing resolution than the 16~ns digitizer sample size, and is free of bias from PMT pulse height.  Fig.~\ref{fig:TOF} shows the relative timing of the gamma, D-T neutron and D-D neutron beams.  The energies of the D-D and D-T neutron beams are calculated from the fitted time differences, $\Delta t$, between the neutron and gamma peaks: 
\begin{equation}
    E_{k} = \frac{1}{2} m_n \left(\frac{1}{\Delta t/L -1/c}\right)^2 .
\end{equation}
The uncertainty in the ToF energy is due to the timing resolution of the digitizer, and stochastic energy loss of deuteron inside the target. 
 
\subsection{Deuteron-Neutron Energy Modeling}
\begin{figure}
\centering
\includegraphics[width=\textwidth]{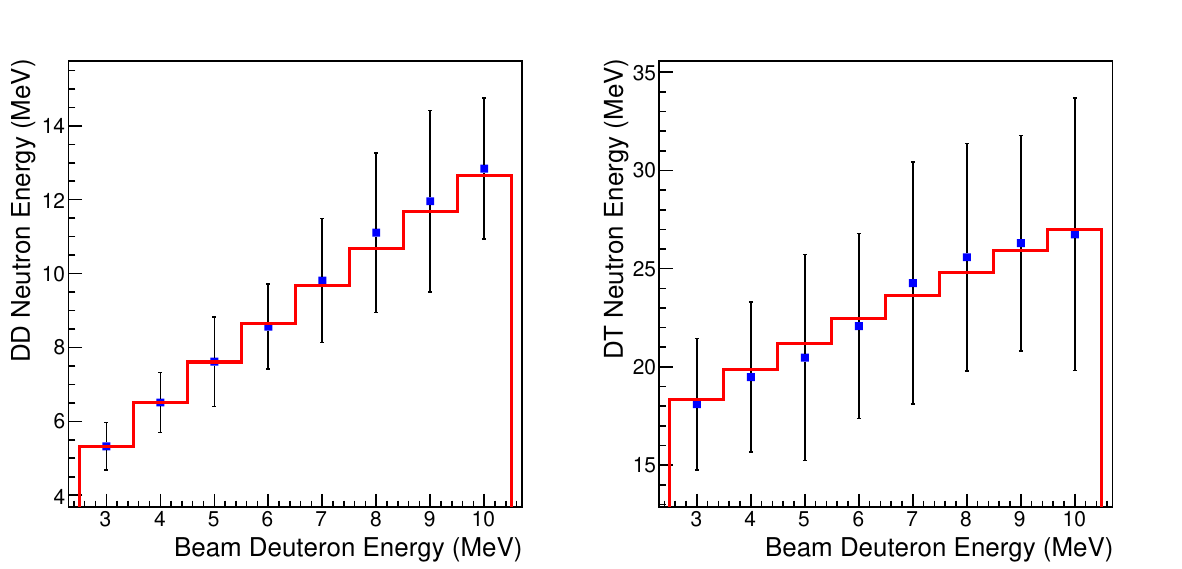}
\caption{The ToF energy (blue dots with error) and best-fit SRIM model prediction of neutron energy (red line).  The thickness of the Havar foil is a fit parameter in the simultaneous fit to the D-D (left) and D-T (right) neutron data. }
\label{fig:srim_dE_model}
\end{figure}

The mono-energetic deuteron beam is accelerated and transported in vacuum, but when it reaches the target there is some energy loss in the Havar foil beam window and helium chamber. The average energy loss was calculated with the SRIM software package~\citep{Ziegler:2010srim}.  The average deuteron energy loss is an integral over dE/dx, as a function of the energy.  From this calculation a 12.5~$\mu$m Havar foil thickness yields the best fit to ToF data (see Fig. \ref{fig:srim_dE_model}), which is available from the manufacturer.  The helium is found to have negligible effect on the deuteron energy.

The neutron energy from the D-T and D-D reactions can be derived by invoking energy-momentum conservation with their respective Q-values, which are 17.59~MeV for D-T and 3.27~MeV for D-D\@~\citep{csikai1987crc}.  With the detector placed directly downstream of the tritiated target, the resulting energy of the forward neutron beam is determined by the deuteron beam energy and Q-value of the reaction. This energy is denoted as the neutron beam energy, $E_n$, which is shown in Tab.~\ref{tab:nsource}. 
The maximum angular deviation from the beam line center due to the geometry of the MicroCHANDLER detector corresponds to a maximum deficit of 2\% in neutron energy.

\subsection{Energy Reconstruction}
Proton recoils induced by the D-D and D-T neutrons are easily separable with a ToF cut.  In the detector, events are reconstructed with the energy reconstruction algorithm developed for MiniCHANDLER, which have been described in detail elsewhere~\citep{Haghighat:2018mve}.  Only minor optimizations are required for the smaller MicroCHANDLER detector.  
The reconstruction efficiency is 99\%, an improvement on the previous 93\% in MiniCHANDLER.  This is due to a lower DAQ zero-suppression threshold and better light collection efficiency of the new PMTs and light guides.

\section{Quenching Factor Calculation}
\begin{figure}[h]
\centering
\includegraphics[width=0.8\linewidth]{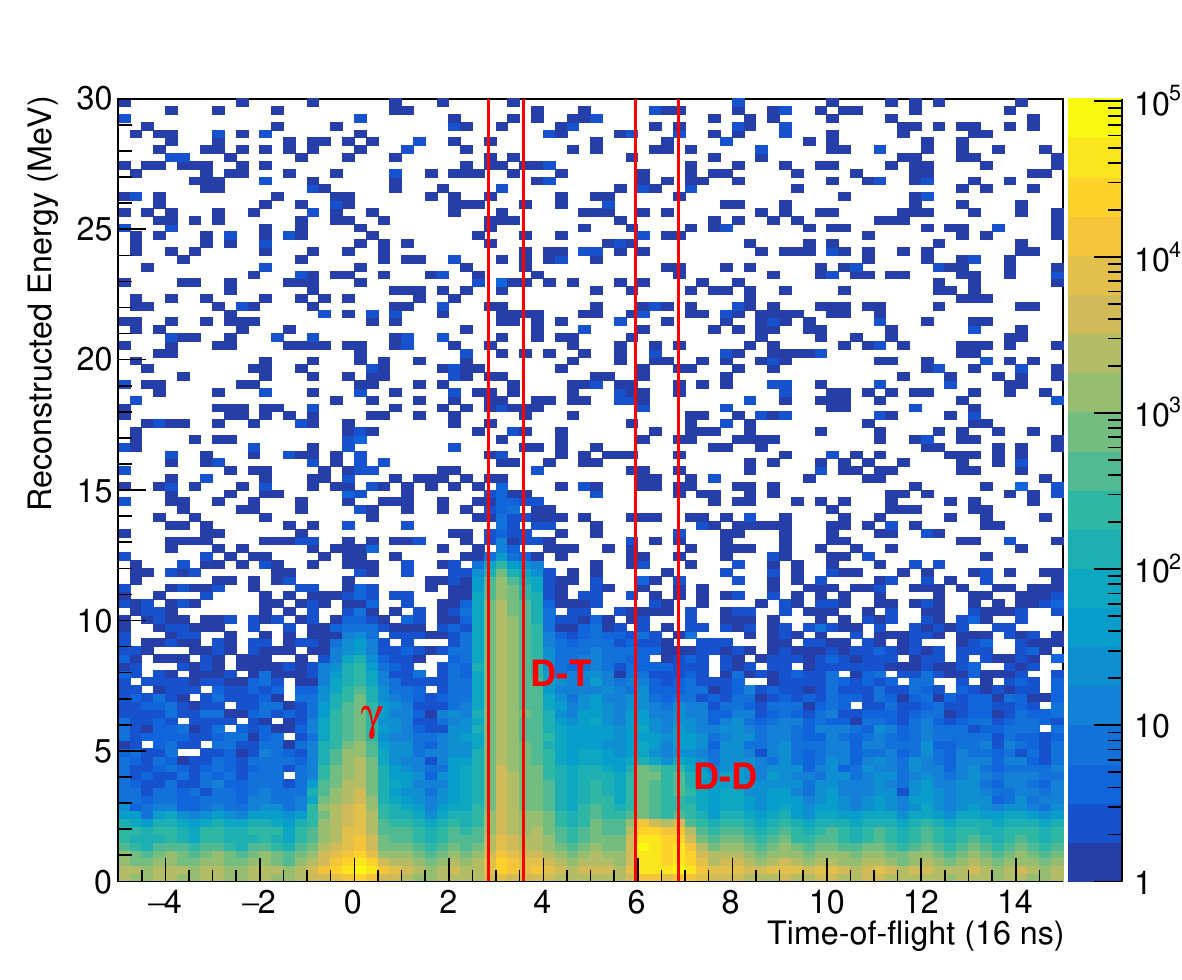}
\caption{2D distribution of ToF vs. reconstructed energy for a terminal voltage of 1.5~MV.  The X-axis is aligned with the gamma peak at t=0. The D-T time window is defined by the 1st and 2nd vertical red lines, while the D-D time window is defined by the 3rd and 4th lines.}
\label{fig:2D}
\end{figure}

 The reconstructed energy spectra from D-D and D-T neutrons were selected by ToF cuts, which are defined to be within $\pm1\sigma$ of the best fit mean ToF for each population, as shown in Fig.~\ref{fig:2D}.    
 This demonstrates a clean separation between the two populations of neutrons.  The y-axis shows a clear difference in the maximum proton recoil energy of the two time windows. 

Two complementary methods are used to extract the proton quenching factor: the end-point fit and the spectrum fit. 

\subsection{End-point Fitting}\label{subsec:ep_fit}
\begin{figure}
\centering
\begin{subfigure}{0.49\textwidth}
\centering
    \includegraphics[width=\textwidth]{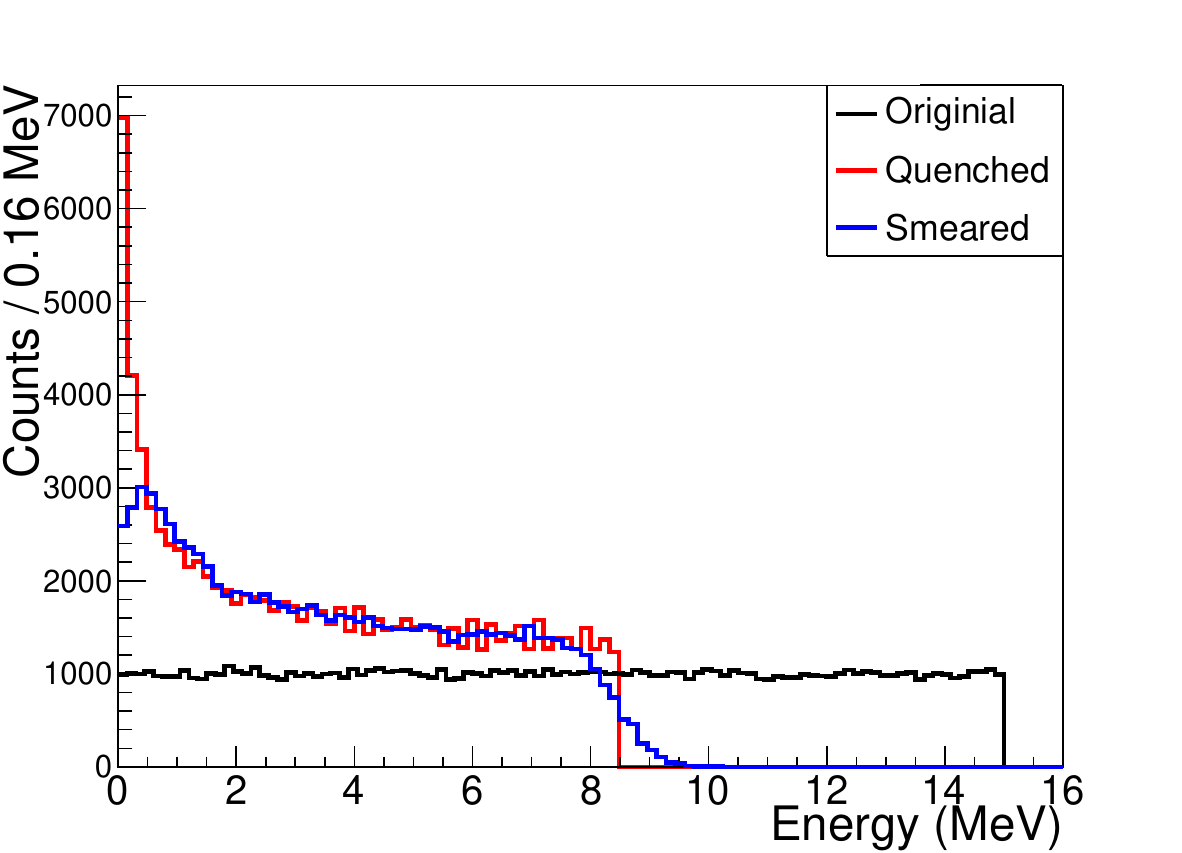}
\end{subfigure}%
\begin{subfigure}{0.49\textwidth}
\centering
    \includegraphics[width=\textwidth]{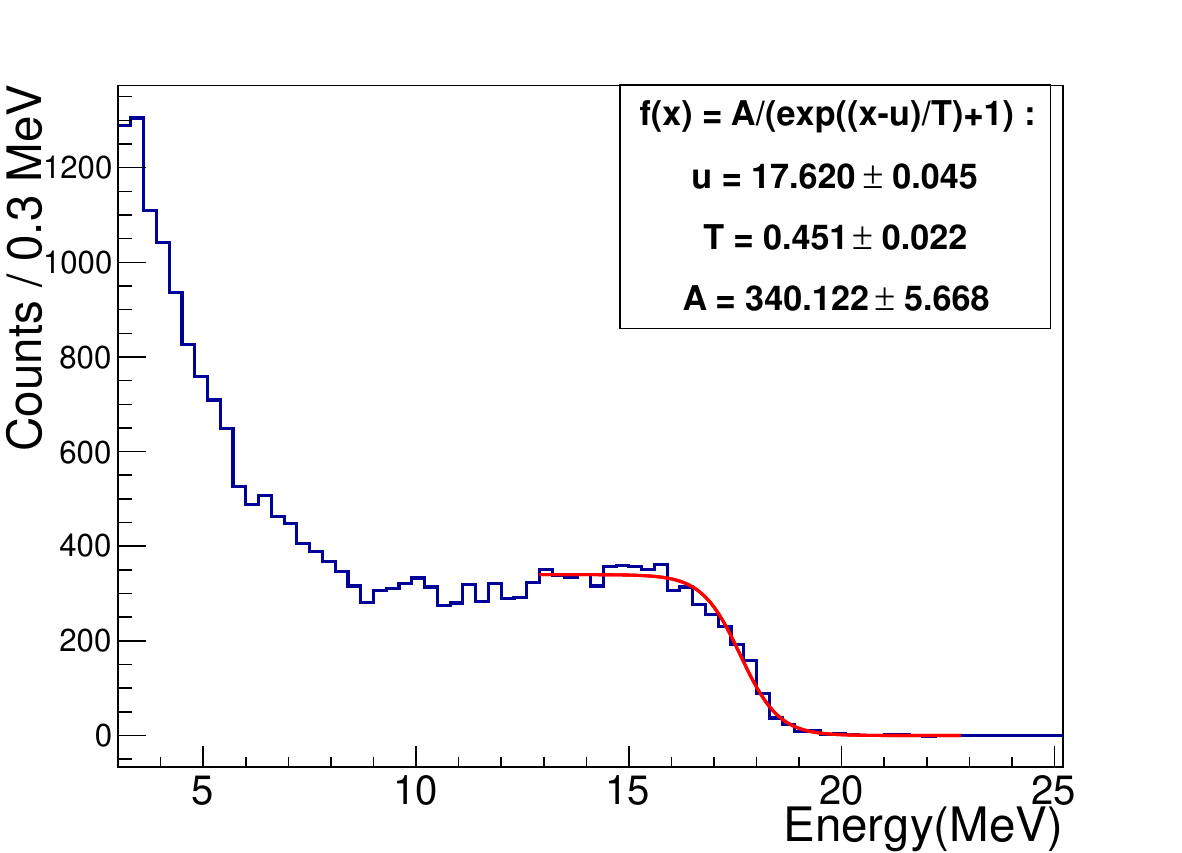}
   
\end{subfigure}
 \label{fig:E_p_data}
\begin{picture}(0,0)
   \put (-395,51) {(a)}
   \put (-182,51) {(b)}
\end{picture}
    \caption{\label{fig:E_p_mock}
    Plot (a) shows the true spectrum (black) from Eq.~(\ref{eq:1scatter}) from a single scattering of a 15~MeV neutron. The quenched spectrum (red) is calculated according to the non-linear relation in Eqs.~(\ref{eqn:birks2}) and (\ref{eqn:birks3}) with realistic quenching factors.  The smeared spectrum (blue) comes from applying a Gaussian smearing to the quenched spectrum, which mimics the detector energy resolution. 
    (b) The visible recoil spectrum (blue) from 25.90\,MeV D-T neutrons is fitted with the function in Eq.~(\ref{eq:fit_func}) (red). The fitted end-point energy is given by parameter $\mathbf{\mu}$.
    }
    \label{fig:E_p}
\end{figure}
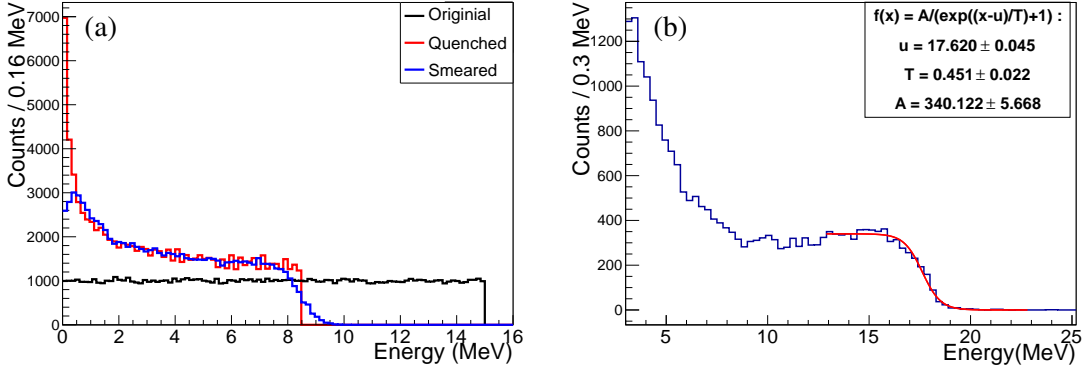

In the end-point fitting method only the maximum edge of each proton recoil spectrum was considered.  Beam neutrons scatter off protons quasi-elastically in the detector.  If only one scatter is considered, the proton recoil energy, $E_p$ is given by:
\begin{equation}
    E_p = \frac{2A(1-\cos\theta)}{(A+1)^2}E_{n,0},
    \label{eq:1scatter}
\end{equation}
where $E_{n,0}$ is the incident neutron energy, $\theta$ is the neutron scattering angle with respect to its initial direction, and $A = \frac{m_p}{m_n} 
\sim 1$.  Therefore, $E_p$ is uniformly distributed from 0 to $E_{n,0}$, with maximum when $\theta=\pi$, and minimum when $\theta=0$.  For the N successive scattering events, the average neutron energy decreases exponentially: 
\begin{equation}
    E_{n,N} 
    \simeq \left(\frac{1}{2}\right)^{\!N}\!\times E_{n,0}.
    \label{eq:nE}
\end{equation}
Hence, by fitting the maximum end-point in the $E_p$ spectrum, complicated calculations at lower energies, due to multiple scattering, are avoided.  The visible proton recoil spectrum also includes the effects of proton quenching and energy smearing, as demonstrated with the mock data study shown in Fig.~\ref{fig:E_p_mock}a.  Here, the change from ``true" to ``quenched" energy shows the impact of proton quenching.  The spectrum end-point shifts from 15~MeV to 8.5~MeV.  The pile-up at lower energies comes from the decreasing quenching factor curve as the proton energy decreases. The final ``smeared" spectrum is obtained with a Gaussian filter on the ``quenched" spectrum. This replicates the features in the data shown in Fig.~\ref{fig:E_p_mock}b. 
The high end of the visible proton recoil energy is fitted with a modified Fermi-Dirac function:
\begin{equation}
    f(E_{vis}) = \frac{A}{e^{(E_{vis}-\mu)/T}+1},
    \label{eq:fit_func}
\end{equation}
where $A$ controls the amplitude of the function, $T$ controls the smearing, and $\mu$ is the quenched end-point, or $E_{ep}$. 
The fit is restricted to the high end of the spectrum (as shown in Fig.~\ref{fig:E_p_mock}b) where multiple recoils make only a tiny contribution.

\begin{figure}[h]
\centering
\includegraphics[width=0.99\linewidth]{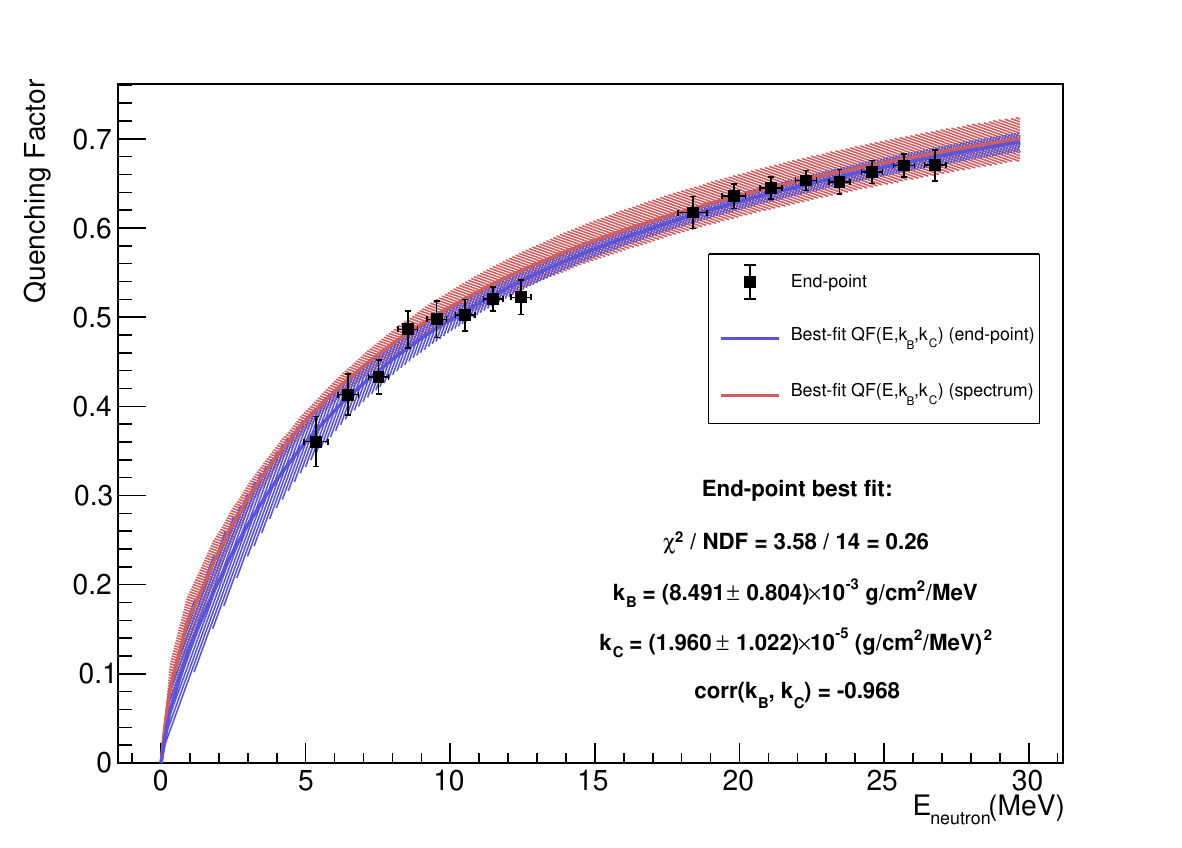}
\caption{Proton quenching factor vs.\ neutron ToF energy.  The proton quenching best-fit curve (blue line) is fitted to the 16 data points (black squares) from the end-point fitting method.  The lower (higher) 8 data points are associated with D-D (D-T) neutrons.  Two functions of $QF(E,k_B,k_C)$ determined by the fitted parameters $k_B$ and $k_C$ are plotted with their 2$\sigma$ confidence bands. The blue curve comes from the end-point fitting method in Sec.~\ref{subsec:ep_fit}, and the red curve comes from the spectrum fitting method in Sec.~\ref{subsec:spec_fit}.  }
\label{fig:QF_fit}
\end{figure}

The proton quenching factors are extracted from the data at 16 distinct energies.  These measured quenching factors $q(E_{n})=E_{ep}/E_{n}$, are plotted against the neutron beam time-of-flight energy, $E_{n}$, in Fig.~\ref{fig:QF_fit}.  A $\chi^2$ function is constructed to fit for values of Birks' constants, $k_B$ and $k_C$:
\begin{equation}
    \chi^2=\sum_{i=1}^{16}\frac{(QF(E^i_n,k_B,k_C)-q^i(E^i_n))^2}{\sigma^2(q^i)+[\sigma(E^i_n)\cdot(\frac{dQF}{dE})_{E=E^i_n}]^2},
    \label{eqn:chi2}
\end{equation}
where $E^i_n$ is the $i$th neutron beam energy, and $q^i$ is the measured quenching factor from each end-point fitting.  The function $QF(E_{n}^i,k_B,k_C)$ is used to fit the quenching factor, which is calculated through a numerical integration of Eq.~(\ref{eqn:birks3})\@.  

Each data point of $q(E_n)$ is determined by a Gaussian mixture of the values and uncertainties from two independent measurements.  The first method uses the full data set, while the second method selects only proton recoil events that are contained in a single cube.  This mixture incorporates the systematic errors in data selection and end-point fitting into the model.  The final $\sigma(q)$ and $\sigma(E_n)$ also include the statistical uncertainty. $dQF/dE$ is derived from the fitted $QF(E,k_B,k_C)$ function to include the uncertainty in $E_{n}$.

The best-fit Birks' constant $k_B = (8.49 \pm 0.80)\times 10^{-3}$~g/cm$^2$/MeV and the second-order parameter $k_C = (1.96 \pm 1.02) \times 10^{-5}\ {\rm (g/cm^2/MeV)^2}$ are extracted using the ROOT TMinuit package~\citep{james1998minuit} with a goodness of fit, $\chi^2_{min}/ndf=3.58/14$.  A test of the quenching factor characterized solely by $k_B$ is made by fixing $k_C$ at zero.  Enforcing this additional constraint increases the best fit in Eq.(~\ref{eqn:chi2}) by $\Delta\chi^2/ndf = 4.73/1$ (a two-sided p-value of  0.0298).  A constraint of $k_B=0$ in the fit yields $\Delta\chi^2/ndf = 84.87/1$.  The parameters $k_B$ and $k_C$ are highly anti-correlated in the fit, with correlation factor of -0.968.  The best-fit quenching factor function is shown by the blue curve in Fig.~\ref{fig:QF_fit}.   A 2$\sigma$ confidence band (blue striped) is calculated using fitted uncertainties derived from a toy MC calculation.  

\subsection{Spectrum fitting}\label{subsec:spec_fit}
An independent spectrum fitting method was also used to evaluate the Birks' constants. In this method, the full spectrum is used to yield higher statistical power.  It also provides a direct test of our MC model, with a comparison across the full energy spectrum. 

\subsubsection{Quenching Spectrum Generation}
\label{sec:quench-spectrum}
The expected quenched proton recoil spectrum is calculated using Eq.~(\ref{eqn:birks2}) and the MCNP true energy depositions.  A raster scan of $k_B$, $k_C$ and a third parameter, $\sigma_E$, is performed to find best fit values. The energy resolution parameter $\sigma_E$ is given by:
\begin{equation}
\label{eqn:smear}
\frac{\sigma_E}{E}=\sqrt{\alpha^2+\frac{\beta^2}{E}+\frac{\gamma^2}{E^2}},
\end{equation}
where $\alpha$ is contributed from the light transmission process, $\beta$ is from the stochastic electron cascade process in the PMTs, and $\gamma$ is from the electronics noise. The contribution from $\gamma$ is found to be negligible.  For each individual proton recoil, $i$, with energy deposition, $E^i$, the quenched energies are calculated and summed to get the total visible energy $E_{sum}$:
\begin{equation}
E_{sum} = \sum_{i=1}^N QF(k_B,k_C,E^i)\cdot E^i,
\end{equation}
where N is the total number of proton recoils in the event, and $QF(k_B,k_C,E^i)$ is calculated from Eq.~(\ref{eqn:birks3}). The speed of the calculation is improved with a pre-calculated table of 300 uniform sample points from 0 to 30 MeV generated for each $k_B,k_C$ pair, and the intermediate values are derived via a linear interpolation.  A Gaussian filter was applied to the binned quenched energy spectrum of $E_{sum}$ to simulate the energy smearing.

\subsubsection{Spectrum Comparison}
Each MC spectrum is compared to the corresponding data spectrum using a modified Pearson's $\chi^2$ test~\citep{Gagunashvili:2006nva}. The goodness of the fit is evaluated through the reduced $\chi^2$ ($\chi^2_r = \chi^2/ndf$).  For simplicity, both MC and data are restricted to the ``single-cube" event sample.

A peak in the data spectrum is present at around 4~MeV corresponding to the 4.4~MeV de-excitation gamma line of the first excited state of $^{12}$C. The de-excitation gammas from $^{12}{\rm C}(n,n)^{12}{\rm C}^*$ inelastic scattering are emitted in the same time window as the proton recoil events and are a prominent feature in the lower energy D-D neutron spectra.  The gamma Compton spectrum is obtained from the MCNP simulation and smeared with the same $\sigma_E$ as the proton recoil spectrum.  In the D-D neutron spectral fits, the 4.4~MeV gamma is fit with an additional nuisance parameter, scaling the simulated spectrum.
\begin{figure}[t]
\centering
\begin{subfigure}{0.49\textwidth}
    \includegraphics[width=\textwidth]{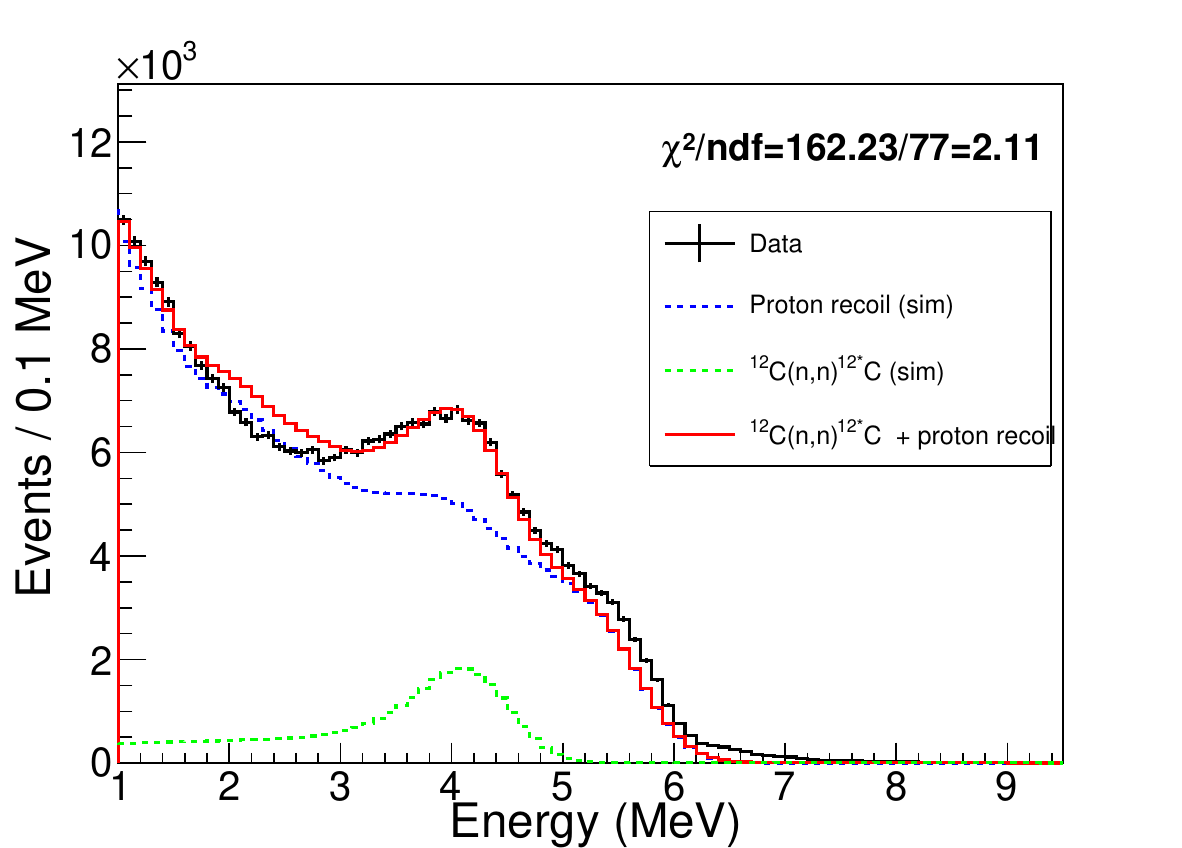}
\end{subfigure}%
\begin{subfigure}{0.49\textwidth}
    \includegraphics[width=\textwidth]{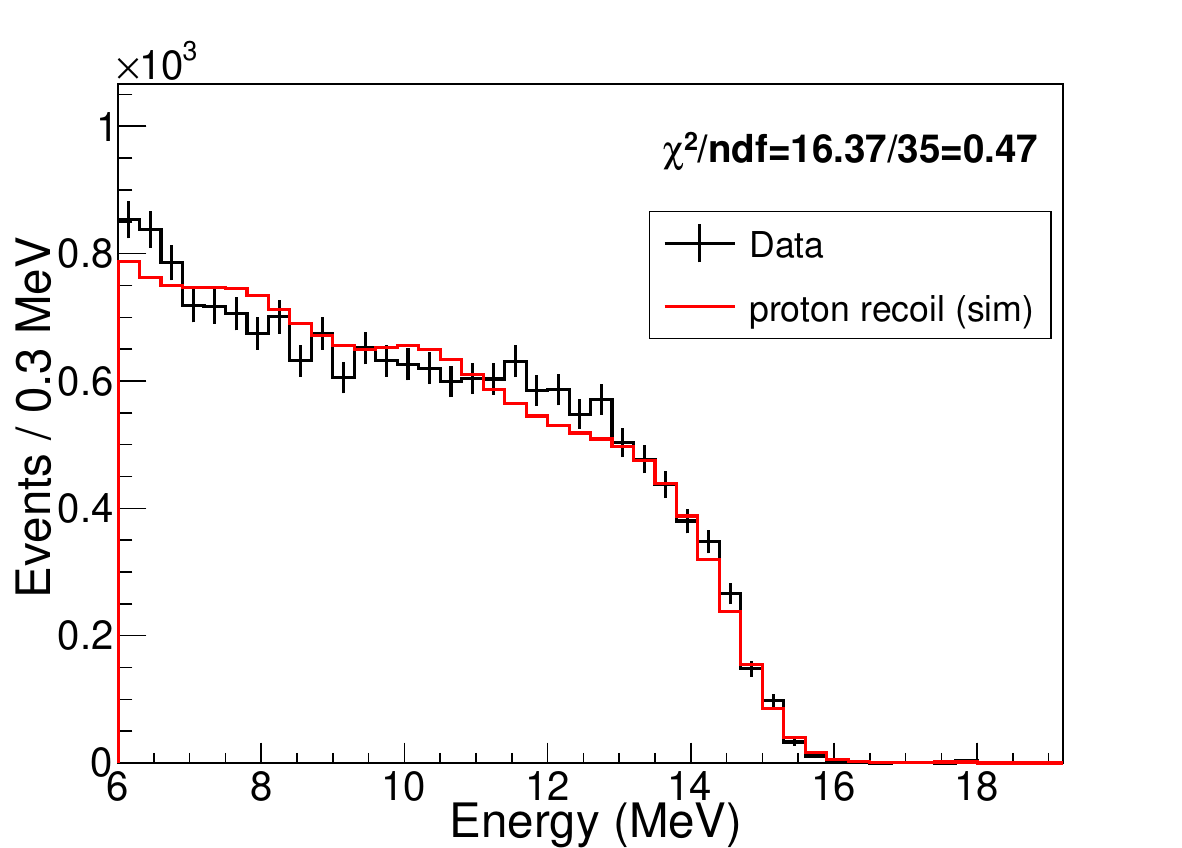}
\end{subfigure}\\
\begin{subfigure}{0.49\textwidth}
    \includegraphics[width=\textwidth]{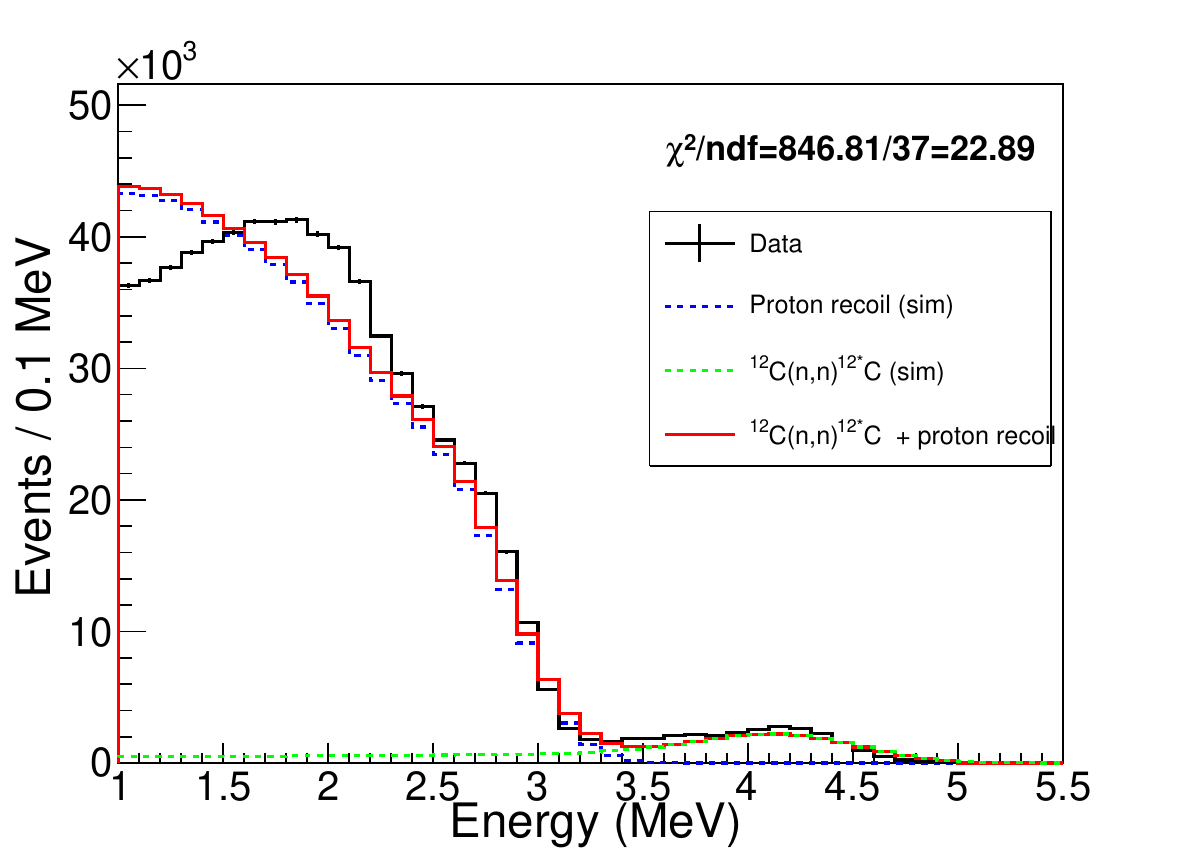}
\end{subfigure}%
\begin{subfigure}{0.49\textwidth}
    \includegraphics[width=\textwidth]{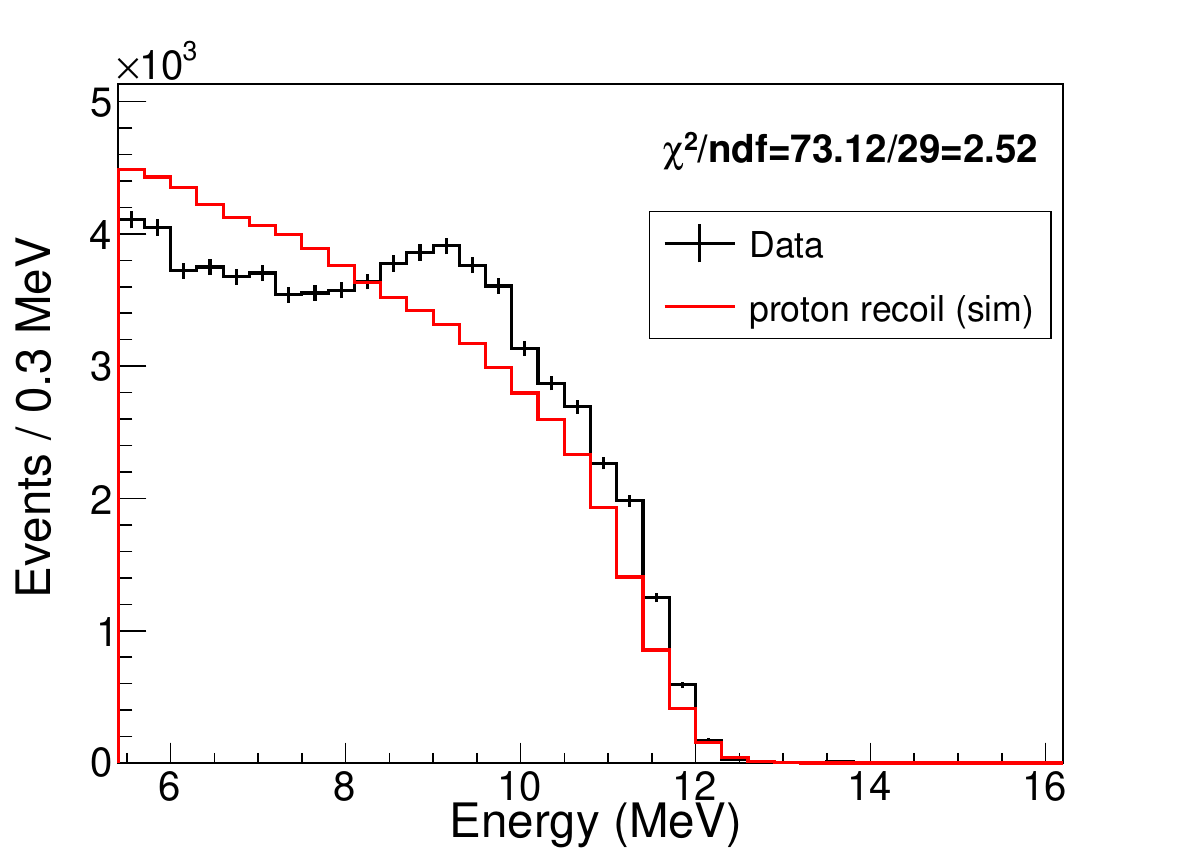}
\end{subfigure}
\begin{picture}(0,0)
   \put (-395,205) {(a)}
   \put (-186,205) {(b)}
   \put (-395,53) {(c)}
   \put (-186,53) {(d)}
\end{picture}
    \caption{Examples of good (a and b) and bad (c and d) spectral fit from the simultaneous fit to all D-D (a and c) and all D-T (b and d) spectra.  D-T recoil spectra are fitted to the proton recoil spectrum from simulation.  Each D-D fit has an additional term for the $^{12}$C de-excitation gamma spectrum.
    }
    \label{fig:spec_fit}
\end{figure}
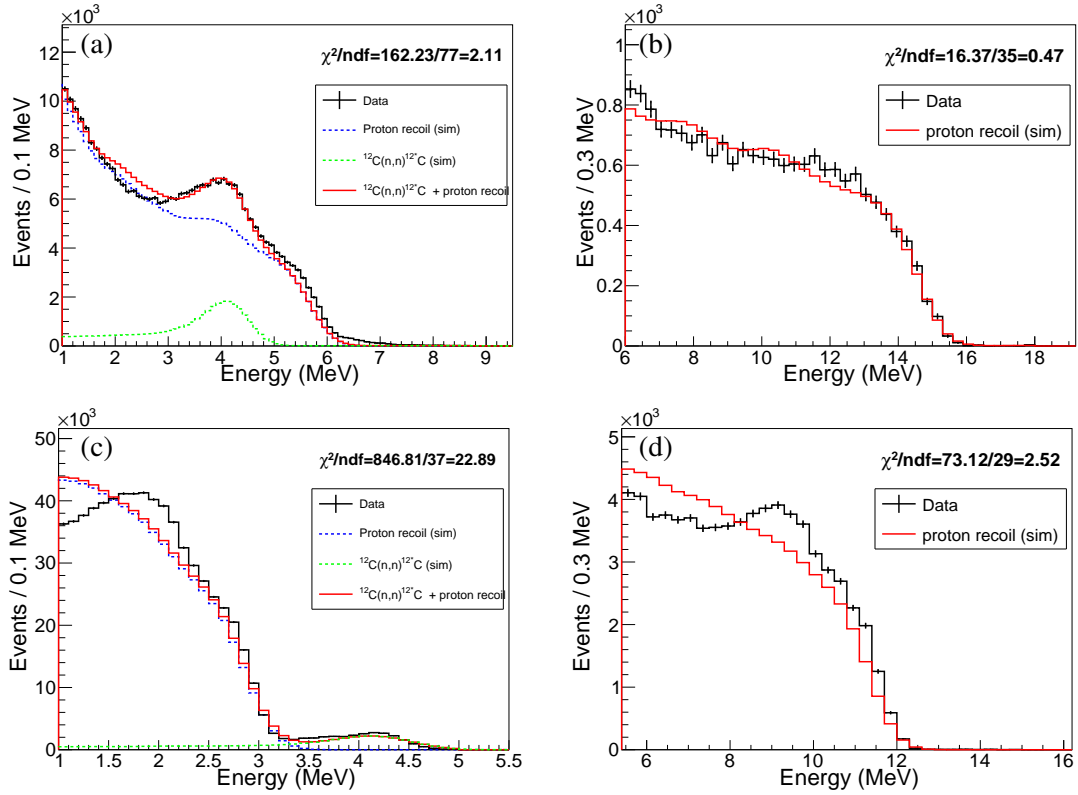

Multiple raster scans of the data to MC $\chi^2$, are done over a 30-by-30 grid in varying regions of interest in ($k_B,k_C$). 20 values of $\sigma_E$ are tested for each combination.  The best fit $\sigma_E/E$ is found to be 5\%/$\sqrt{E\,{\rm(MeV)}}$ ($\alpha=0,\beta=0.05$) for both D-D and D-T neutron fits.  The best-fit results in the end-point analysis are adopted as the baseline value, denoted by $k_B^0$ and $k_C^0$.

\begin{figure}[h]
\centering
\includegraphics[width=0.8\linewidth]{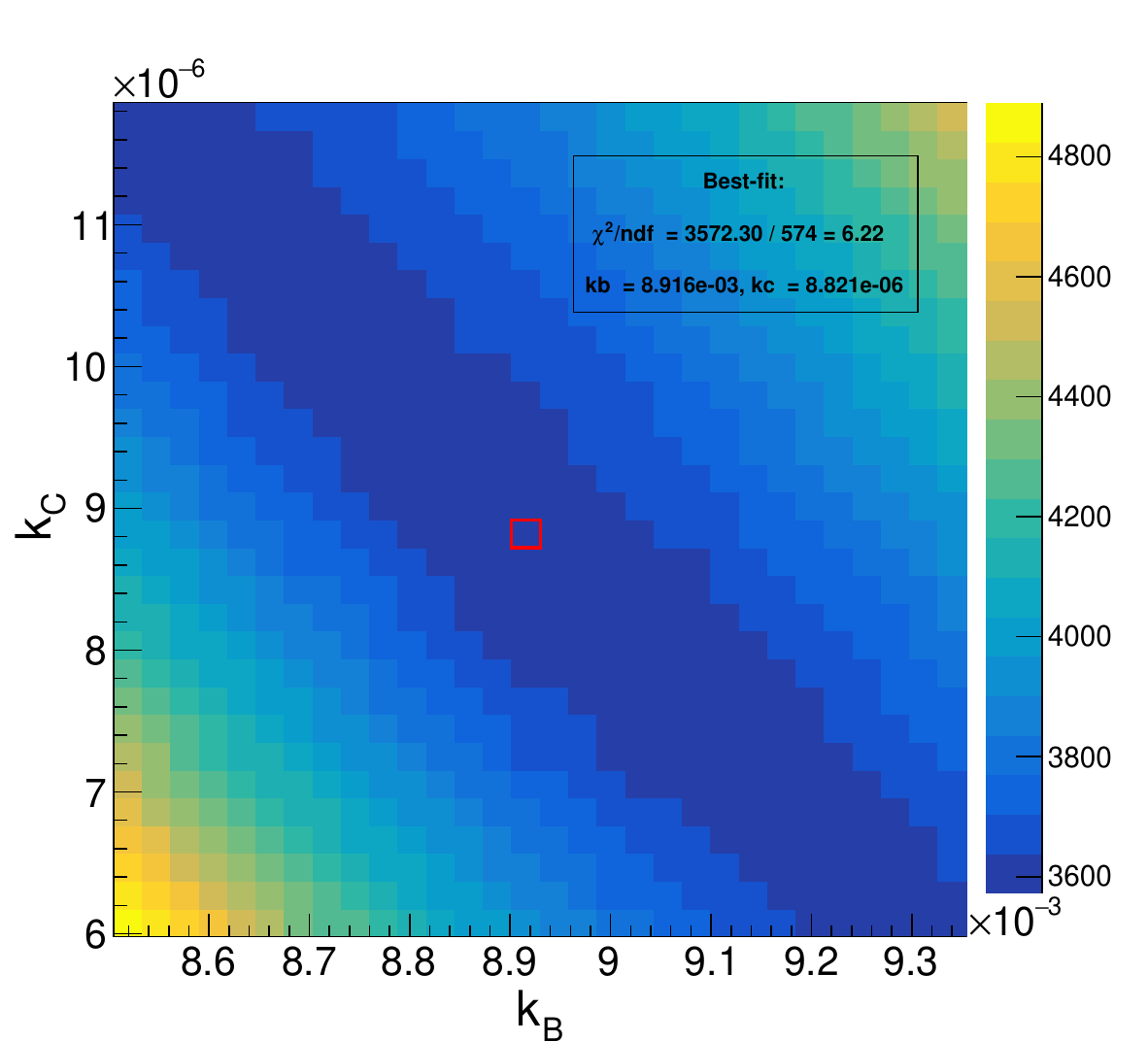}
\caption{Raster scan of $\sum \chi^2$ as a function of $k_B,k_C$. The lowest $\chi^2$ grid point is denoted by the red box.}
\label{fig:kb_kc_scan}
\end{figure}

In performing a $\chi^2$ minimization between data and MC on the eight D-D neutron spectra only, a $\chi^2/ndf=3193/554=5.76$ is found, with the best-fit $(k_B,k_C) = (1.01 k_B^0, 0.56 k_C^0)$.  Similarly, fitting the eight D-T neutron spectra gives $\chi^2/ndf=319/276=1.16$, with the best-fit $(k_B,k_C) = (1.09 k_B^0, 0.32 k_C^0)$.  Fig.~\ref{fig:spec_fit} shows examples of both good and bad spectral fits from these simultaneous fits.  By evaluating each individual fitted spectrum, the lowest two D-D neutron spectra were found to have the largest contributions to the total $\chi^2$.  In these two spectra, there are fewer events in lower energy region, which is likely due to the ADC threshold of the detector.  The combined D-D and D-T fit yields the best-fit $(k_B,k_C) = (1.05 k_B^0, 0.45 k_C^0)$ with $\chi^2/ndf=3572/832=4.29$.  A distribution of the total $\chi^2$ over the the 2D raster scan space is shown in Fig.~\ref{fig:kb_kc_scan}.

The statistical uncertainty of the combined D-D and D-T fit is derived by varying $k_B$ and $k_C$ centered at their best-fit values, until the increments of $\chi^2$ correspond to 1 sigma. 

Here, a 0.6\% statistical uncertainty is estimated for $k_B$, and a 5.2\% statistical uncertainty is estimated for $k_C$.
The systematic uncertainty from combined spectrum fit is estimated by performing individual ``local" fits to each of the 16 spectra, and evaluating the fluctuations in each best-fit parameter.  This uncertainty is found to be 11.2\% and 69.2\% for $k_B$ and $k_C$, respectively. 

The systematic due to the uncertainty of the input ToF energy in MC is not modeled as iterations with full-chain MC are computationally heavy.  This will be covered in a future MC study on MicroCHANDLER\@.  The final best-fit Birks' constants from the spectrum fit method are $k_B = (8.92 \pm 1.00)\times 10^{-3}\ {\rm g/cm^2/MeV}$ and $k_C = (8.82 \pm 6.10) \times 10^{-6}\ {\rm (g/cm^2/MeV)^2}$.  The $QF(E,k_B,k_C)$ function corresponding to these values is shown as the red curve in Fig.~\ref{fig:QF_fit}, and the 95\% confidence band is given by the red striped region. 

\subsection{Discussion}
Fig.~\ref{fig:QF_fit} shows the results from the end-point fitting method and the MC spectrum fitting method are in agreement within their error bands.  In hypothesis testing, the Birks' constants derived from these two methods, $k_B$ and $k_C$, both agree to within 1$\sigma$. Therefore, the results from the two methods are consistent. 

\begin{figure}[h]
\centering
\includegraphics[width=0.99\linewidth]{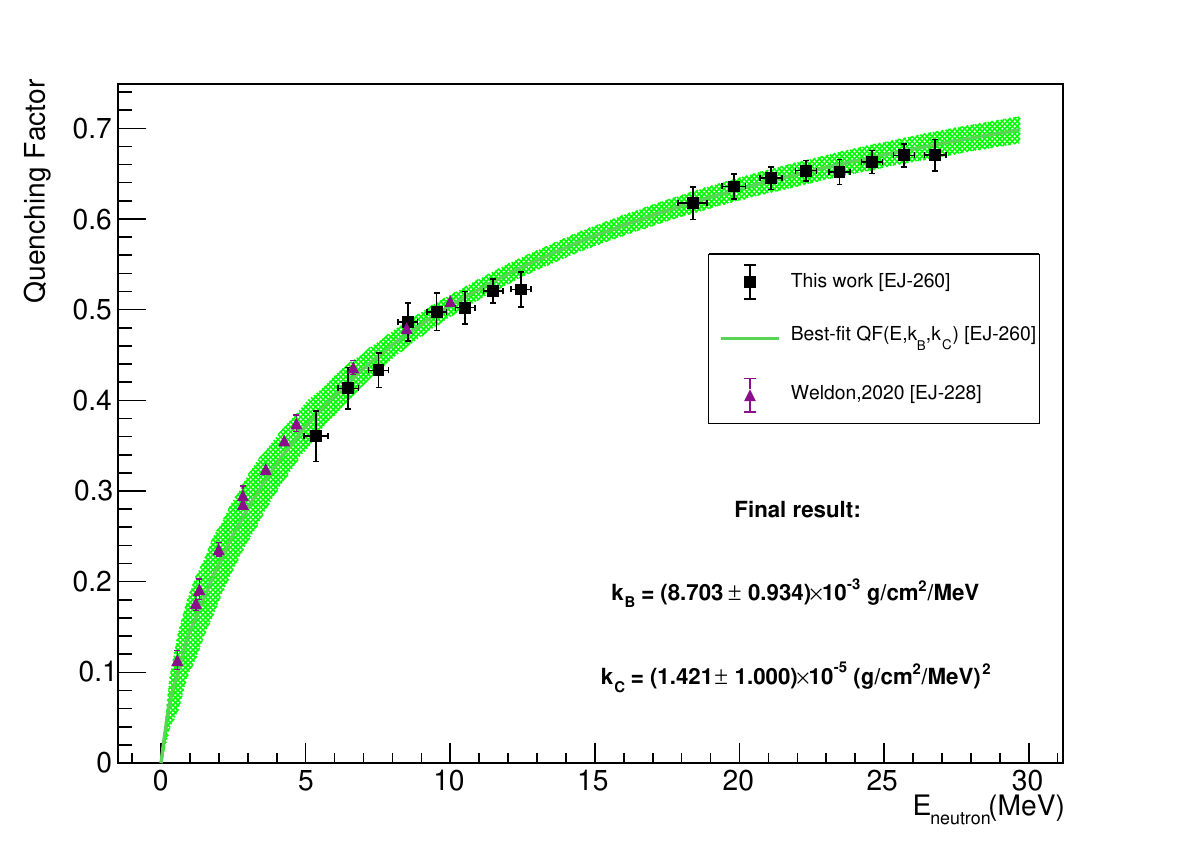}
\caption{ Proton quenching factor vs.\ neutron ToF energy.  The function of $QF(E,k_B,k_C)$ with its 2$\sigma$ confidence bands from the combined analysis of end-point fitting and MC spectrum fitting.  The purple triangles are from an earlier proton light yield measurement~\citep{Weldon:2020} on a similar scintillator. }
\label{fig:QF_fit_combined}
\end{figure}

The final result is reported as the Gaussian mixture of the two results, where $k_B = (8.70 \pm 0.93)\times 10^{-3}\ {\rm g/cm^2/MeV}$ and $k_C = (1.42 \pm 1.00) \times 10^{-5}\ {\rm (g/cm^2/MeV)^2}$.  The $QF(E,k_B,k_C)$ function and the 95\% confidence band for this combined result are shown in Fig.~\ref{fig:QF_fit_combined}. A proton light yield measurement on a similar scintillator, covering lower energies~\citep{Weldon:2020} is reproduced in the same plot for reference. 

\section{Conclusion}
The proton quenching of EJ-260 plastic scintillator was measured in the MicroCHANDLER detector.  A collimated deuteron beam on TiT target was used to generate neutron beams with energies from 5 to 27~MeV through D-D and D-T interactions. Sixteen neutron beam energies were used.   Using reconstructed energy spectra measured in the MicroCHANDLER detector, the 16 quenched proton light yields were fit to extract the Birks' constants with high precision from quenching factors.  The visible energies covered by this result span the region of interest for fast neutron backgrounds in a reactor IBD detector.  A detailed Monte Carlo model was constructed in MCNP to predict the  proton's response in the detector.  Covering a unique energy range, this measurement will benefit the modeling of energy non-linearity in plastic scintillator detectors, and become a valuable input to simulating fast neutron backgrounds in future surface-level detectors like CHANDLER.

\acknowledgments

This work was supported in part by the National Science Foundation, under grant number IIP-1924433; Virginia Tech's Institute for Critical Technology and Applied Science; U.S. Department of Energy under grant numbers DE-FG02-97ER41033 and DE-SC0020235; and the U.S. Department of Energy, National Nuclear Security Administration, Office of Defense Nuclear Nonproliferation R\&D through the consortium for Monitoring, Technology and Verification under award number DE-NA0003920.


\bibliography{mybibfile}

\end{document}